# Modeling of Coal Drying before Pyrolysis

Damintode Kolani[1, a], Eric Blond[1, b], Alain Gasser[1], Tatiana Rozhkova[2], Matthieu Landreau[2]

[1]PRISME Laboratory - University of Orléans,
Polytech Orléans, 8, rue L. de Vinci, 45072 Orléans, France

[2]Centre de Pyrolyse de Marienau (CPM),
rue Moulins, 57600 Forbach, France

[a]damintode.kolani@univ-orleans.fr, [b]eric.blond@univ-orleans.fr

**Abstract.** The coking process is composed of two main stages: drying process and pyrolysis of coal. A heat and mass transfer model was developed to simulate the drying process of coal. The mechanisms of heat and mass transfer described in the model are: conduction through the coal cake; conduction and convection through the gas in pores; generation, flux and condensation of water vapor. The model has been implemented in finite element software. It requires basic data on the coke oven charge properties and oven dimensions as input. These input data were obtained by experiments or from the literature. The proposed model includes condensation and evaporation allowing us to reproduce the temperature plateau observed experimentally.

**Introduction**

The coking process consists to eliminate the volatile matter from coal by distillation in the absence of air and at atmospheric pressure. During the coking process, there are two main stages: the dry phase allows the vaporizing moisture contained in the coal without chemical decomposition and pyrolysis converts dry coal into gas and coke [1]. The final aim of the European project SPRITCO which includes this work is the prediction of the wall pressure on the heating walls in the coke oven battery to prevent their damage and then to increase their lifetime. The model presented deals with the first stage, the coal drying. The drying process is a complex mechanism involving simultaneous phenomena of mass transport, heat transfer and volume change. Several studies have been performed over the last two decades to develop models to simulate the drying kinetics [2-8]. Considering the scale chosen to write the model and the driving force taken into account, two main approaches are used to characterize the mass transfer phenomena during the drying process: the overall macroscopic approach and the microscopic one. The macroscopic approach comprises two sub approaches: multi-components and single component approaches.

The multi-components approach describes separately the different components constituting the system: porous material, water and air. Indeed, for each component of the system (water, air, water vapor), mechanisms of mass and heat transfer are described independently. Several authors have adopted this approach, which was initially proposed by Whitaker [9] to describe the process of mass and heat transfer in porous media. To complete the formulation, the constitutive equations for each phase are integrated over a representative elementary volume of the porous medium. The obtained equations of conservation of energy and mass are applicable at the macroscopic scale for each phase. Whitaker's approach has been applied by several authors [2-6, 10-11]. Models based on this approach require the determination of several physical parameters, which makes them particularly complex.

The single component approach is an energetic approach which considers the gradient of water potential as the sole driving force responsible for the water movement in the porous medium. Luikov [7-8] has developed a set of coupled partial differential equations to describe heat and mass transfer in porous media assuming that moisture transfer is analogous to the heat transfer and the mass transfer is proportional to the gradient of moisture and temperature. Its model was based on the Onsager reciprocal relations of the thermodynamic of irreversible processes, giving a set of equations of transfer with the phenomenological coefficients which are functions of the porous medium. Two main groups of models derived from the approach of Luikov, the diffusion models [12-13] and the models based on the energetic potential [14-16].

The microscopic approach consists to link up the anatomical structure to the macroscopic parameters required for modeling [17-19]. The goal is to build a link between the different scales, of microscopic observation until the drying process. This allows developing drying models based on properties at a microscopic scale.

The model presented in this paper is based on a multi-components macroscopic approach. This modeling method overcomes the modeling difficulty induced by the heterogeneity of the porous medium. This model does not take into account the mechanical behavior of coal cake. So, the coal shrinkage induced by drying is not considered.

**Mathematical modeling**

Considering the granular structure of coal, the presence of moisture and volatile matter, the coal charge is modeled as a partially saturated porous medium. A porous medium is defined as a material volume consisting of a continuous solid matrix with interconnected voids filled by a mixture of dry air, water vapor and liquid water. The volume fractions are described by the porosity $\emptyset$ and the saturation $S$. The porosity is the ratio of pore volume to total volume and the saturation is the ratio of liquid water volume to pore volume. So, the volume fraction of solid is $1 - \emptyset$ while the volume fraction of liquid is $S\emptyset$. In a complementary way, the volume fraction of gas is defined by $(1 - S)\emptyset$.

In order to develop the model, a continuum approach is adopted and representative elementary volume is considered [20]. The binary gas mixture is assumed to follow the ideal gas behavior. So, gas molar densities are described by:

$$\rho_i = \frac{P_i M_i}{RT} \quad \text{for } i = v, a \tag{1}$$

where $P_i$ is the partial pressure of constituent i, $M_i$ is the molar mass, R is the universal gas constant, equal to 8.314 J mol$^{-1}$ K$^{-1}$, T is the temperature, $v$ is water vapor and $a$ is dry air. The molar mass of gas phase is given by [21]:

$$\frac{1}{M_g} = \frac{w_v}{M_v} + \frac{(1-w_v)}{M_a} \quad \text{with} \quad w_i = \frac{\rho_i}{\rho_g} \tag{2}$$

where $w$ is the mass fraction and $g$ is the gas phase.

The process of coal drying involves simultaneous transport of heat and mass through the coal cake, a three-phase porous medium. Transfer of heat and moisture in a rigid and unsaturated porous medium involves interaction of two different physics (heat and mass transfer) and three different phases (solid, liquid and gas). Therefore, for this problem, the following assumptions are introduced: local thermal and phase equilibrium exist between the three phases at every point in the coal cake and an inert porous solid is assumed, so that the solid density remains constant. Liquid water is assumed immobile, capillary and gravity effects are neglected. Indeed, the liquid water phase may not be continuous and so the liquid displacement due to Darcy law and gravity can be neglected.

For the calculation of relative velocity, the Lagrangian frame with reference the skeleton is chosen. Vapor transport mechanisms result from gas mixture global movement and diffusion in the gas. The velocity $V_g$ of the gas mixture is governed by Darcy's law:

$$\vec{V}_g = -\frac{k}{\mu_g} \overrightarrow{grad}(P_g) \tag{3}$$

where $k$ is the intrinsic permeability (m$^2$), $\mu_g$ is the dynamic viscosity (Pa s) of gas and $P_g$ is the total pressure of the gas. The diffusive flux of each gas constituent is governed by Fick's law:

$$\vec{J}_i = -\rho_g D_{eff} \overrightarrow{grad}(w_i) \tag{4}$$

where $D_{eff}$ is the effective diffusion coefficient (m$^2$ s$^{-1}$) which depends on temperature and pressure. It is given by [20]:

$$D_{eff} = D_{eff}^{ref} \left(\frac{T}{T^{ref}}\right)^{1.88} \frac{P_{atm}}{P_g} \tag{5}$$

where $D_{eff}^{ref}$ is the reference effective diffusion coefficient, $T^{ref}$ is the reference temperature, $P_{atm}$ is the atmospheric pressure.

The mathematical formulation of the problem is a set of partial differential equations that combines mass balance, energy balance and state laws of each phase.

The general local form of the mass balance of a species $j$ is [5, 20, 22]:

$$\frac{\partial \bar{\rho}_j}{\partial t} + \text{div}(\rho_j \vec{V}_j) = \dot{m}_j \tag{6}$$

where $\bar{\rho}_j$ is the apparent average density, $\rho_j$ is the intrinsic mass density, $V_j$ is the velocity of phase j relative to the solid skeleton and $\dot{m}_j$ is the mass rate due to phase change. For the mass conservation of liquid water phase, the source term is the mass rate of evaporation. It is obtained thanks to the non-equilibrium approach [6, 23-25]:

$$\dot{m}_{evap} = \frac{KM_v}{RT}(P_{sat} - P_v) \tag{7}$$

where K is a penalty factor and $P_{sat}$ is the saturated vapor pressure given by the Clausius-Clapeyron equation:

$$P_{sat} = P_{atm} \exp\left(\frac{M_w \Delta H_{evap}}{R}\left(\frac{1}{T_o} - \frac{1}{T}\right)\right) \tag{8}$$

where $M_w$ is the water molar mass, $\Delta H_{evap}$ is the latent heat of evaporation and $T_o$ is the boiling point. The saturated vapor pressure depends only on temperature, and is independent of the air pressure and moisture.

To avoid evaporation when no liquid is available, the penalty factor is expressed as follows:

$$K = K_o\left(1 - H(S^r - S)\right) \tag{9}$$

where $K_o$ is the inverse of the time step and $S^r$ is the residual saturation.

For immobile liquid water, mass balance of liquid water phase yields:

$$\frac{\partial \bar{\rho}_w}{\partial t} = -\dot{m}_{evap} \tag{10}$$

where $\bar{\rho}_w$ is the apparent density of liquid water.

The mass balance of gas phase is**:**

$$\frac{\partial \bar{\rho}_g}{\partial t} + \text{div}(\rho_g \vec{V}_g) = \dot{m}_{evap} \tag{12}$$

The mass flux of the gas mixture could be linked to the flux of vapor and dry air:

$$\rho_g \vec{V}_g = \rho_v \vec{V}_v + \rho_a \vec{V}_a \tag{13}$$

The mass flux of a constituent of gas mixture is given by:

$$\rho_i \vec{V}_i = \rho_i \vec{V}_g + \vec{J}_i \tag{14}$$

The transport of fluids through coal cake can be subdivided into two main categories: firstly, gas inside the material moves by convection due to a pressure gradient and secondly, water vapor migrates via a diffusion mechanism.

The mass balance of each gas constituent is:

$$\frac{\partial \overline{\rho_i}}{\partial t} + \text{div}(\rho_i \vec{V}_g + \vec{J}_i) = \dot{m}_i \tag{15}$$

where $\dot{m}_j$, the mass rate due to phase change varies as function of gas constituent. So, for water vapor, it corresponds to mass rate of evaporation but for dry air, it is null because there is no production of dry air during the drying process.

The heat balance accounting for the diffusive and convective heat transfer in the whole medium linked to gas mixture flow is expressed as follows [21, 26]:

$$\overline{\rho C_P}\frac{\partial T}{\partial t} + \rho_g C_{p,g} \vec{V}_g \cdot \overrightarrow{\text{grad}}(T) + \text{div}\left(\lambda_{eff}\overrightarrow{\text{grad}}(T)\right) = \dot{Q} \tag{16}$$

where

$\overline{\rho C_P}$ is the overall volumetric specific heat given by:

$$\overline{\rho C_P} = \rho_{app} C_{p,eff} \tag{17}$$

where $\rho_{app}$, the equivalent mass density is calculated by means of the weighted sum of mass density of different phases and $C_{p,eff}$, the effective specific heat is obtained experimentally [27].

The convective term is given by:

$$\rho_g C_{p,g} \vec{V}_g = \rho_v C_{p,v} \vec{V}_v + \rho_a C_{p,a} \vec{V}_a \tag{18}$$

The effective thermal conductivity $\lambda_{eff}$ is also obtained experimentally.

The source term is given by:

$$\dot{Q} = -\dot{m}_{evap}\Delta H_{evap} \tag{19}$$

The first term represents the average heat capacity of the environment (energy accumulation), the second term corresponds to the heat transport by convection and diffusion due to gas transport. Finally, the third term expresses the heat conduction represented by the effective conductivity of the medium, the last one is the source term which takes into account evaporation.

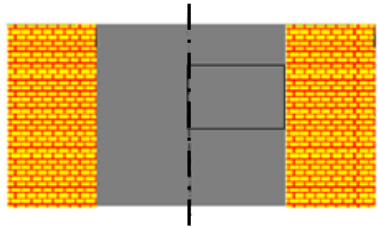

Fig. 1: Geometric configuration

The coke oven used is that of Centre de Pyrolyse de Marienau. Fig.1 represents the geometric configuration of the model: at both sides, the heating walls and at the middle, the coal cake. The oven dimensions are 465mm in width, 1000mm in length and 1050mm in height. Taken into consideration the symmetry of the charge and the load, only one half could be simulated. The simulation is achieved under the assumption of the homogeneity of coal cake. This has allowed the simplification of the simulation by taking only a slice of one half of the geometry as shown in Fig. 1 where the black rectangle represents the computation area.

For the gas mixture, the initial gas pressure is specified as the atmospheric pressure. The ambient temperature is the initial temperature. Depending on the composition of the coal and the moisture content, the initial saturation and initial vapor mass fraction are estimated.

The boundary condition at the center line of the oven is set as a symmetry boundary condition. At the surface of the coal in contact with heating wall, the heat flux is specified including convection and radiation. The inward heat flux is:

$$\vec{n}.\left(\lambda_{eff}\vec{\nabla}T\right) = h_T(T - T_b) + \sigma_0\epsilon(T^4 - T_b^4) \qquad (20)$$

where $\vec{n}$ is the outward normal vector, $h_T$ is the heat convection coefficient, $\sigma_0$ is the Stefan-Boltzmann constant, $\epsilon$ is the surface emissivity and $T_b$ is the heating wall temperature which is equal to 1000°C.

The transport of each gas constituent through the boundary is specified by the flux of the constituent and is expressed as follows:

$$\vec{n}.\left(w_j\,\rho_g\vec{V}_g + \vec{J}_j\right) = h_m(\rho_j - \rho_j^b) \qquad (21)$$

where $h_m$ is the mass transfer coefficient and $\rho_j^b$ is the constituent mass density at wall temperature.

**Results and discussion**

In this present model, the coal used is a low volatile coal. It contains 7.2 wt% of moisture, 18.8 wt% of volatile matter and 6.4 wt % ash on dry basis and its bulk density on dry basis is $738\,\text{kg}\,\text{m}^{-3}$. Its elemental analysis on dry basis is 86 wt% C, 4.23% H, 1.12 wt % O, 1.47 wt % N, 0.78 wt % S. During the drying process, thermal properties of coal vary as a function of temperature. The permeability variation as function of temperature is obtained experimentally. Between the room temperature and the boiling temperature, its magnitude order is $10^{-12}$. In these calculations, the parameters used are listed in Table 1.

Table 1: Values of parameters used in the proposed model

| Parameter | Symbol | Value | Unit |
|---|---|---|---|
| heat convection coefficient | $h_T$ | 10 | $W.m^{-2}.K^{-1}$ |
| surface emissivity | $\epsilon$ | 0.9 | - |
| mass transfer coefficient | $h_m$ | $10^{-6}$ | $m.s^{-1}$ |
| Porosity | $\emptyset$ | 33.9 | % |
| initial saturation | $S_o$ | 11 | % |
| heating wall temperature | $T_b$ | 1000 | °C |
| factor penalty | K | 0.1 | $s^{-1}$ |

The vapor flux over the thickness is shown in Fig. 2 (a). At x=0, the coke oven centre and x=232.5 mm, the face of coal cake in contact with the coke oven. As the temperature increases gradually from the outside towards the middle of the charge, vapor moves into both directions as seen from the sign of the vapor flux. A positive flux means transport toward the heating wall and a negative flux indicates transport towards the centre of the coal cake. So, at the beginning, both evaporation and condensation occur. After 5h, only evaporation is observed because the saturated vapor pressure is greater than the partial vapor pressure at any point of the charge. Fig. 2 (b) shows the change of saturation as function of time. Gradually as the temperature increases, evaporation occurs, then saturation decreases gradually until it becomes null. During evaporation, a part of the vapor migrates through the coal cake by convection and diffusion, and condenses in colder regions towards the centre. Evaporation and condensation of water are governed by the equilibrium between the partial vapor pressure and the vapor saturation pressure. Condensation takes place when the partial vapor pressure exceeds the saturation pressure while evaporation occurs when the partial vapor pressure is below the saturation pressure and if liquid water is available.

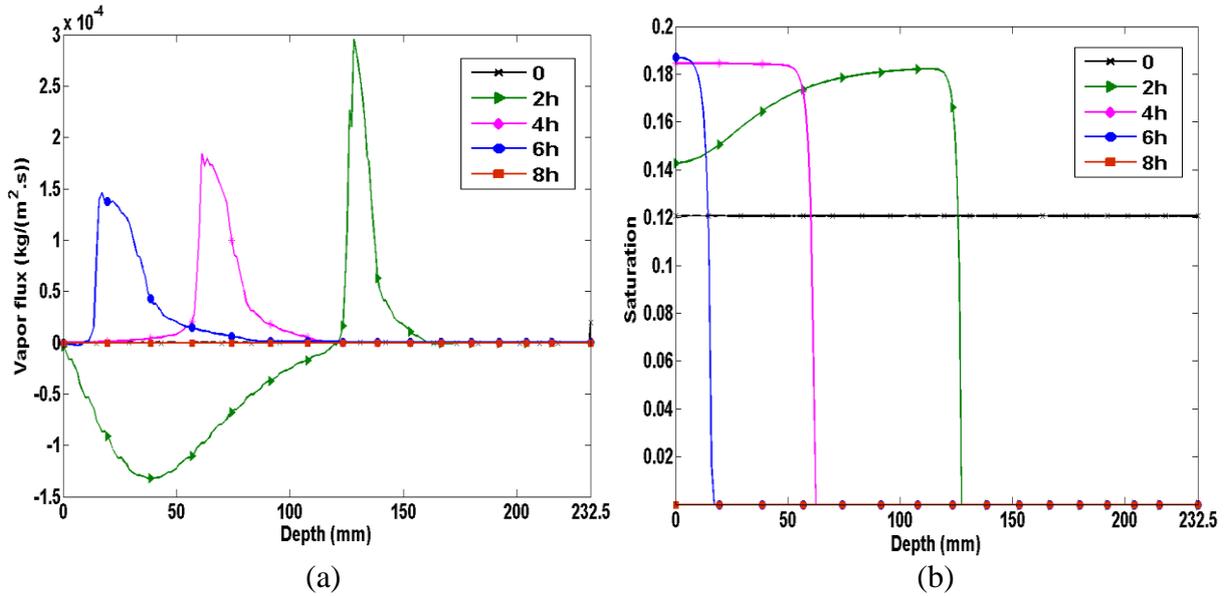

Fig. 2 (a) Vapor flux over the thickness and (b) Saturation as function of time.

Fig. 3 (a) shows the temperature evolution in depth as function of time. The temperature increases gradually from the outside towards the centre. In this figure, at around 100°C, a change of the curve shape is observed. This change is due to the phase change of water, from liquid to vapor, which leads to a temperature plateau at around 100°C, this phenomenon is illustrated in Fig. 3 (b). The end of the temperature plateau is reached faster numerically than experimentally. The main reason is that the curve obtained numerically corresponds to the drying process of coal while the curve obtained experimentally corresponds to the coking process. The latter takes into account all phenomena present during pyrolysis [28]. During pyrolysis, produced tars vapors follow the same path as moisture vapor: they go toward the center and then condense in this colder area. Then, a non negligible amount of heat is used to re-evaporate these tars instead of heating the coal solid skeleton. Besides, there is some additional amount of carbonization water formed in carbonization reactions. The above additional amount of water is governed by the elemental composition of coal. That is why the temperature plateau takes more time during the coking process than during the drying process of coal. Beyond 100°C, the comparison between the curves cannot be done because the drying process is finished. It should be noted that the model is very sensitive to the coal properties especially its thermal properties which are difficult to measure.

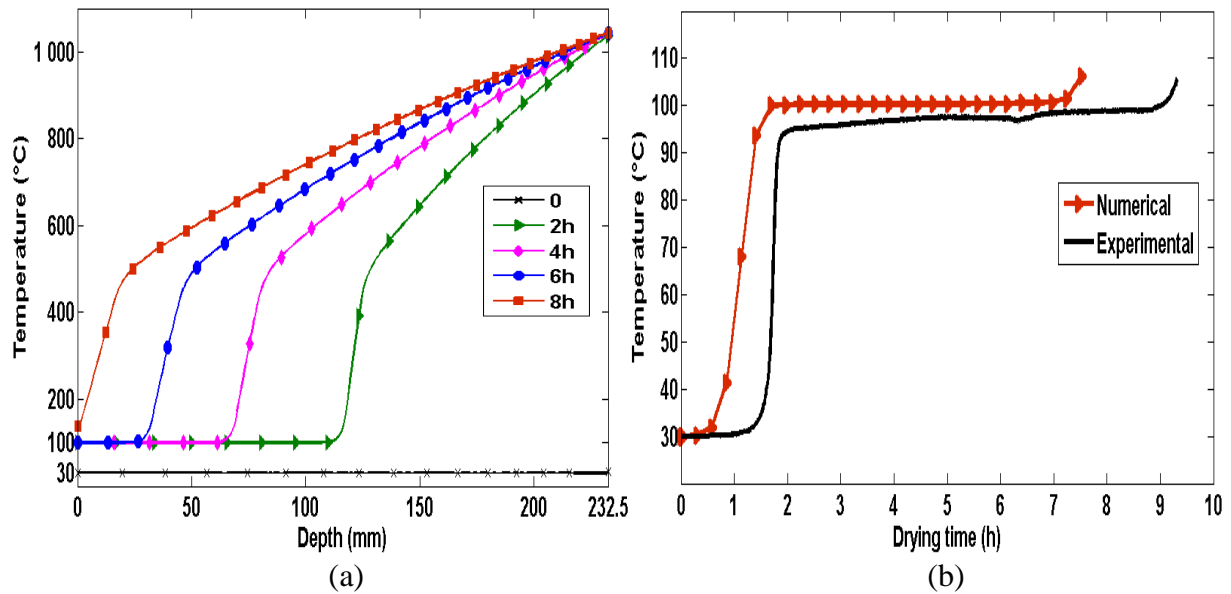

Fig. 3 Temperature evolution as function of time (a) in depth and (b) in centre.

**Conclusions**

This work described heat and mass transfer during coal drying process. The present model reproduced the main mechanisms observed during the process. The simulation result predicts evaporation and condensation of vapor, gas flow paths and temperature profile of the coal cake. Condensation and evaporation have a significant effect and are responsible for the temperature plateau. This work constitutes the first step of coking process modeling. Thereby, an implementation of coal pyrolysis is in progress in order to be able to estimate the wall pressure during this process.

**Acknowledgements**

The authors would like to acknowledge European Union for its financial support for SPRITCO research project of RFCS program, and all partners of this project.